\documentclass[10pt,journal,compsoc]{IEEEtran}
\usepackage{graphicx}
\usepackage{url}
\usepackage{multirow}

\ifCLASSOPTIONcompsoc
  \usepackage[nocompress]{cite}
\else
  \usepackage{cite}
\fi
\ifCLASSINFOpdf
\else
\fi
\usepackage{amsmath}
\usepackage{amsfonts}
\begin{document}
%
\title{Hiding Data Hiding}
%
%
%
%

\author{Hanzhou Wu, Gen Liu and Xinpeng Zhang
\thanks{Hanzhou Wu, Gen Liu and Xinpeng Zhang are with the School of Communication and Information Engineering, Shanghai University, Shanghai 200444, China. E-mail: h.wu.phd@ieee.org, 704105800@qq.com, xzhang@shu.edu.cn. This draft is only used to ensure timely dissemination of the authors' research.}}

%
%

\markboth{}%
{}
%



\IEEEtitleabstractindextext{%
\begin{abstract}
Data hiding is the art of hiding secret data into a cover object such as digital image for covert communication. In this paper, we make the first step towards hiding ``data hiding'', which is totally different from many conventional works that directly embed secret data in a given cover object. In detail, we propose a novel method to disguise data hiding tools, including a data embedding tool and a data extraction tool, as a deep neural network (DNN) with an ordinary task (i.e., style transfer). After training the DNN for both style transfer and data hiding, while the DNN can transfer the style of an image to the target one, it can also hide secret data into a cover image or extract secret data from a stego image. In other words, the tools of data hiding are hidden to avoid arousing suspicion. Experimental results and analysis have shown the feasibility, applicability and superiority of the proposed method.
\end{abstract}

}

\maketitle

\IEEEdisplaynontitleabstractindextext

%
\IEEEpeerreviewmaketitle

\section{Introduction}
\IEEEPARstart{D}{ata} hiding (DH) \cite{journal:dh:ibm, book:mdh, book:stego} enables us to hide secret data into a digital object (typically called \emph{cover}) such as image, video and audio by slightly modifying the noise-like component of the cover. The resultant object carrying secret data (also called \emph{stego}) will not introduce noticeable artifacts and should be sent to the data receiver through a probably insecure channel such as social network and the Internet. By using the secret key for data extraction, the data receiver is able to reconstruct secret data from the stego object. 

Many advanced DH algorithms \cite{journal:emd:cl, journal:codes:zwm, journal:hdf:wang, journal:EA:luo, journal:tcsvt:wu} have been proposed in the literature. A straightforward idea for categorizing DH algorithms is the type of the used cover. For example, image is still the most popular cover source nowadays due to its wide distribution over social networks and ease of editing. With the rapid development of multimedia computing and wireless communication technologies, other covers such as video sequences \cite{journal:tifs:xdw, journal:jvcir:xdw, journal:video:tdsc:chenyi, journal:video:tdsc:chenyanli}, speech signals \cite{journal:speech:access:xue, journal:speech:mtap:ma}, texts \cite{Yang:TIFS, Yang:TIFS2, Kang:MWSF}, and even behaviors \cite{Wu:graph} are attracting more and more attention. Though many DH algorithms can be extended to different covers, there should be unique treatment for each cover because a particular cover always has its own statistical and perceptual characteristics. Therefore, there are two common principles for designing DH systems \cite{book:stego}. One is to preserve the selective model of the cover source during data embedding \cite{stego:mb}. The other one aims to minimize the well-defined distortion introduced by data embedding such as \cite{conference:HUGO:pevny, journal:idf:wang}.

While conventional DH algorithms need much empirical knowledge of the algorithmic designer, recent great success in deep learning (DL) \cite{DL:book} has motivated many researchers to apply DL for DH. For example, increasing works such as \cite{Hidden:Fei-Fei-Li, steganoGAN, LFM:CVPR, ABDH:AAAI:Paper, DH:TPAMI} use deep neural networks (DNNs) for both embedding secret data into a cover and extracting secret data from the stego by an (almost) end-to-end fashion. Following cost based embedding strategy \cite{conference:HUGO:pevny}, mainstream DH works, e.g., \cite{GAN:TIFS:Jianhua, DRL:TIFS:Tang}, use DNNs to learn the optimal (or near-optimal) modification probabilities of cover elements so that a cost function can be derived, which allows a data hider to embed secret data with the minimized distortion.

The aforementioned algorithms expect to make the presence of secret data in the stego statistically undetectable. In real-world applications, the adversary may not only analyze the content, but also monitor the behavior. It implies that the data hider and the data receiver should not only conceal the presence of secret data, but also prevent the data embedding tool and the data extraction tool from being exposed so as to ensure security. In other words, the data hiding tools should be hidden, which is defined as \emph{hiding ``data hiding'' (HDH)}.

Motivated by the above perspective, we propose a HDH method that disguises data hiding tools as a DNN with an ordinary task in this paper. In the proposed method, a DNN is trained with images for data embedding, images for data extraction and images for style transfer. After training, the DNN has the ability to embed secret data into a cover image, extract secret data from a stego image, and transfer the style of a given image to the target one. In other words, the DNN accomplishes three different tasks that are data embedding, data extraction and style transfer. Experimental results and analysis have verified the feasibility, applicability and superiority of the proposed HDH method.

The rest structure of this paper is organized as follows. We first introduce the proposed method in Section 2. Then, in Section 3, we conduct extensive experiments and provide analysis. Finally, we conclude this work in Section 4.

\begin{figure*}[!t]
\centering
\includegraphics[width=6.5in]{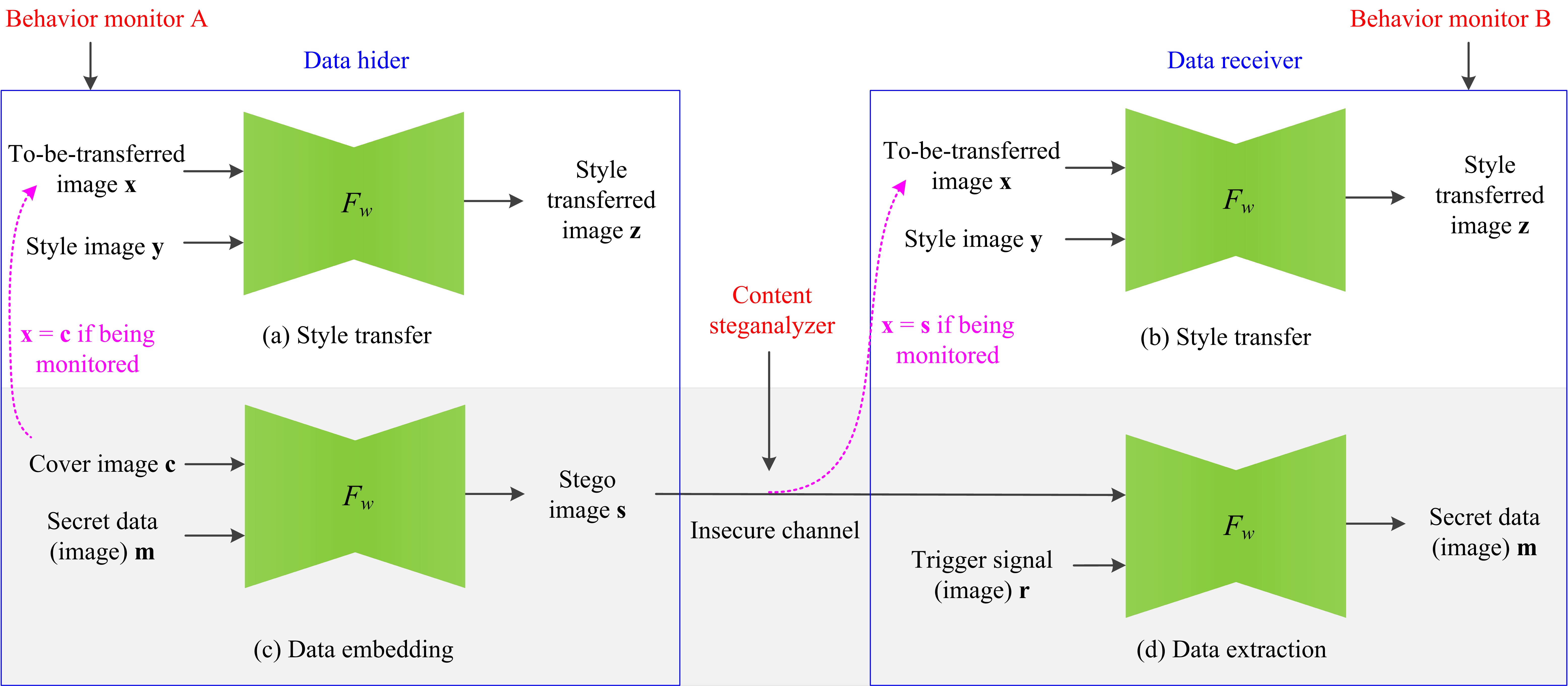}
\caption{An application scenario for the proposed HDH method.
All input data and output data are images. The four DNNs shown here are completely identical to each other including network structure and parameters. ``$\textbf{x}$ = $\textbf{c}$ ($\textbf{s}$) if being monitored'' means to transfer the style of $\textbf{c}$ ($\textbf{s}$). Both the data hider and the data receiver should have the style image $\textbf{y}$ so that they can use $F_w$ for style transfer. The data receiver has the trigger signal $\textbf{r}$.
}
\end{figure*}

\section{Hiding Data Hiding}
\subsection{Overview}
The proposed HDH method aims to produce such a DNN $F_w$ that it can complete three different tasks: \emph{data embedding}, \emph{data extraction} and \emph{style transfer}. The purpose of data embedding is to embed secret data into a cover image to generate a stego image. The purpose of data extraction is to extract secret data from a stego image. For style transfer, it transfers the style of an image to the target one. Mathematically, for style transfer, given a to-be-transferred image $\textbf{x}$ and a style image $\textbf{y}$, $F_w$ generates an image $\textbf{z} = F_w(\textbf{x}, \textbf{y})$, whose style is the same as $\textbf{y}$. For data embedding, given a cover image $\textbf{c}$ and secret data $\textbf{m}$, $F_w$ generates a stego image $\textbf{s} = F_w(\textbf{c}, \textbf{m})$, whose visual quality is close to $\textbf{c}$. For data extraction, given the stego image $\textbf{s}$ and a trigger $\textbf{r}$ (image), $F_w$ reconstructs the secret data $\textbf{m} = F_w(\textbf{s}, \textbf{r}) = F_w(F_w(\textbf{c}, \textbf{m}), \textbf{r})$ from $\textbf{s}$. The style transfer task will be used to conceal the existence of the data embedding task and the data extraction task. 

 For better understanding, we here present an application scenario for the proposed work. As shown in Fig. 1, there are five participants: \emph{data hider}, \emph{data receiver}, \emph{content steganalyzer}, \emph{behavior monitor A} and \emph{behavior monitor B}. The entire process can be described as follows. The data hider wants to share secret data $\textbf{m}$ with the data receiver. If he is being monitored, he feeds a cover image $\textbf{c}$ and a style image $\textbf{y}$ to $F_w$ to generate a style transferred image $\textbf{z}$ to avoid arousing suspicion from the behavior monitor A. Otherwise, he feeds $\textbf{c}$ and the secret data $\textbf{m}$ to $F_w$ to generate a stego image $\textbf{s}$, which will be sent to the data receiver and may be analyzed by the content steganalyzer. After receiving the stego image $\textbf{s}$, the data receiver feeds $\textbf{s}$ and the trigger signal $\textbf{r}$ to $F_w$ to reconstruct the secret data $\textbf{m}$ if the data receiver is not being monitored by the behavior monitor B. Otherwise, the data receiver feeds $\textbf{s}$ and the style image $\textbf{y}$ to $F_w$ for style transfer to avoid arousing suspicion from the behavior monitor B.
 
Both the data hider and the data receiver hold $F_w$, which can be designed and well trained by cooperation in advance. In the following, we introduce the details of $F_w$, including network architecture, input, output, and loss function. 

\subsection{Network Architecture}
It is open for us to design the network architecture of $F_w$. For simplicity, we use a simple but effective DNN, which was inspired by U-Net \cite{dnn:unet}. As shown in Fig. 2, two images with a size of $128\times 128\times 3$ are concatenated to constitute a tensor (with a size of $128\times 128\times 6$) that will be then fed into a down-sampling layer, whose output is a feature map with a size of $64\times 64\times 64$. The output will pass through six down-sampling layers to generate a feature map with a size of $1\times 1\times 512$, which will be then up-sampled to produce a feature map with a size of $2\times 2\times 512$. By concatenating this feature map and the previously generated feature map, another new feature map is generated, which will be processed with five up-sampling layers according to a similar way. Finally, by using a transposed convolutional layer, the final image with a size of $128\times 128\times 3$ can be generated. Fig. 3 shows the details for the up-sampling and down-sampling modules.

Different from U-Net \cite{dnn:unet} that accepts only one image as the input, the network architecture here accepts two images as the input. The reason is that both style transfer and data embedding requires us to provide two images. In order to match the input, a trigger image is used for data extraction. Different from U-Net that uses the max-pooling operation to capture regions of interest for efficient image segmentation, we skip the max-pooling operation since it will cause a lot of useful information to be lost and therefore would impair the data extraction task. Moreover, there has no need to use cropping, which has been used in \cite{dnn:unet}. In addition, we use batch normalization \cite{paper:bn} for accelerating model training and dropout \cite{paper:dropout} for preventing the model from overfitting.

\subsection{Input and Output}
It is necessary to describe the input and output for the three different tasks in detail. For data embedding in Fig. 1 (c), the secret data \textbf{m} is a binary image, which consists of $\frac{128}{N}\times \frac{128}{N}$ disjoint pixel blocks with a size of $N\times N\times 3$. A pixel block can be considered as a sub-image of the entire image. The values of pixels in a block depend on the secret bit to be embedded. Clearly, for each block, if the secret bit to be embedded is ``0'', all values of pixels in the block are set to 0. Otherwise, all values of pixels in the block are set to 1. Since each block carries one bit, the total number of secret bits is $\frac{128}{N}\times \frac{128}{N}$. We define it as the \emph{actual length (AL)} of $\textbf{m}$.

For data extraction in Fig. 1 (d), the trigger signal \textbf{r} is an image activating $F_w$ to extract $\textbf{m}$ from the stego image $\textbf{s}$. It is open for us to design $\textbf{r}$. For simplicity, we use the image whose pixel values are all zero as $\textbf{r}$. $\textbf{r}$ should be used for training $F_w$, which will be detailed later. One thing to note is that each block in the output image of $F_w$ should be mapped to either ``0'' or ``1'' in order to reconstruct the secret bit. This can be done by determining the average value of all elements in the block. Specifically, the last layer of $F_w$ uses tanH, which limits each activation value to the range $[-1, 1]$. For each block, we determine the average value of all elements. If the average value is higher than zero, the secret bit is considered as ``1''. Otherwise, it will be ``0''.

For style transfer in Fig. 1 (a, b), the purpose of the style image $\textbf{y}$ is to ensure that $F_w$ can generate a style transferred image $\textbf{z}$ for the to-be-transferred image $\textbf{x}$. $\textbf{z}$ and $\textbf{y}$ have the same style. It is possible that $\textbf{x} = \textbf{c}$ or $\textbf{x} = \textbf{s}$, which means to transfer the style of $\textbf{c}$ or $\textbf{s}$ (refer to the dotted line in Fig. 1).

\begin{figure}[!t]
\centering
\includegraphics[width=3.2in]{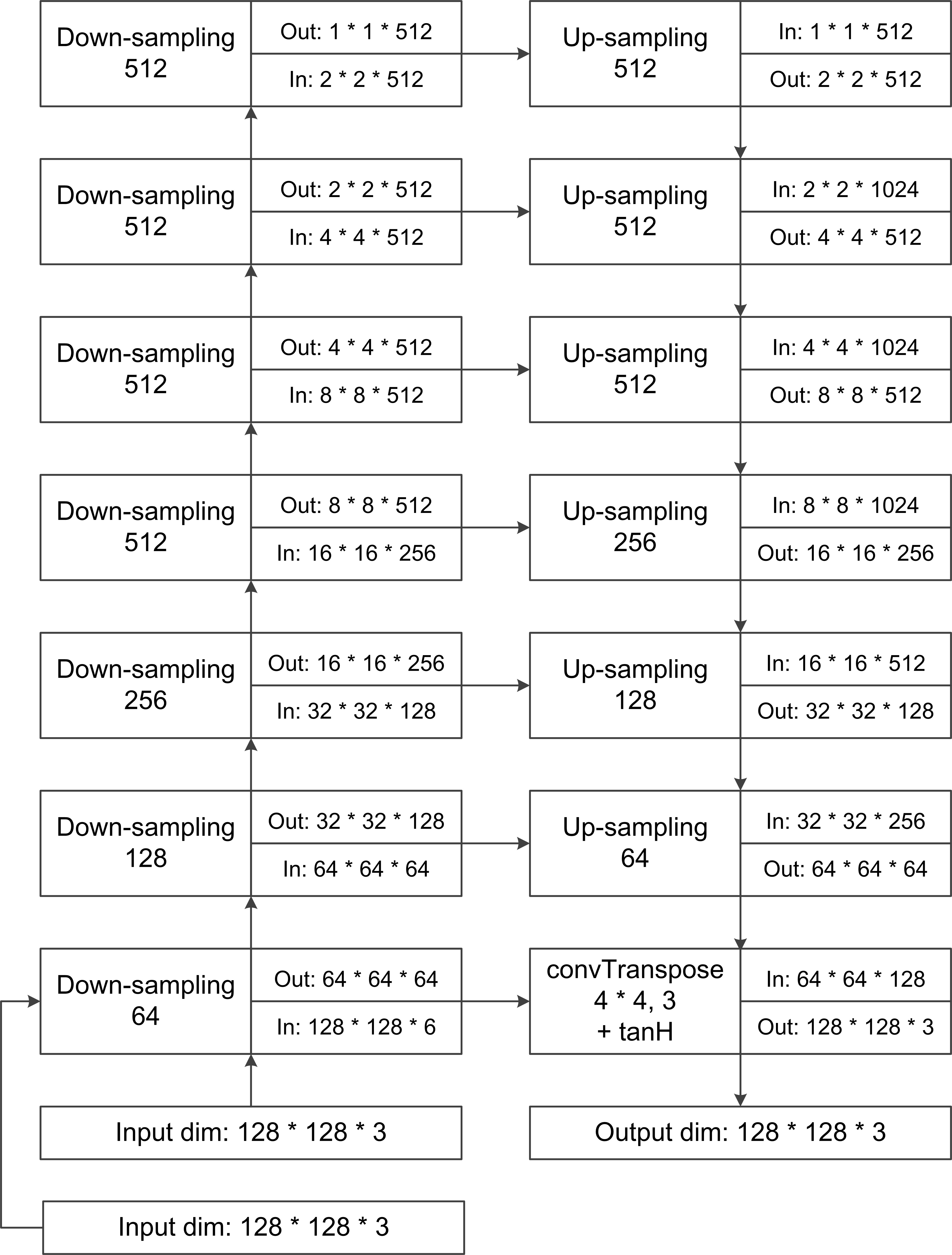}
\caption{Network architecture. All up-sampling layers (including the last layer) use the transposed convolution \cite{trans:conv}. The up-sampling layers use ReLU \cite{relu:leakyrelu} and the down-sampling layers use LeakyReLU \cite{relu:leakyrelu}. All up-sampling layers and all down-sampling layers except for the first down-sampling layer use batch normalization (BN) \cite{paper:bn} so as to accelerate the convergence. The first two up-sampling layers use dropout (= 0.5) \cite{paper:dropout}.}
\end{figure}

\begin{figure}[!t]
\centering
\includegraphics[width=3.2in]{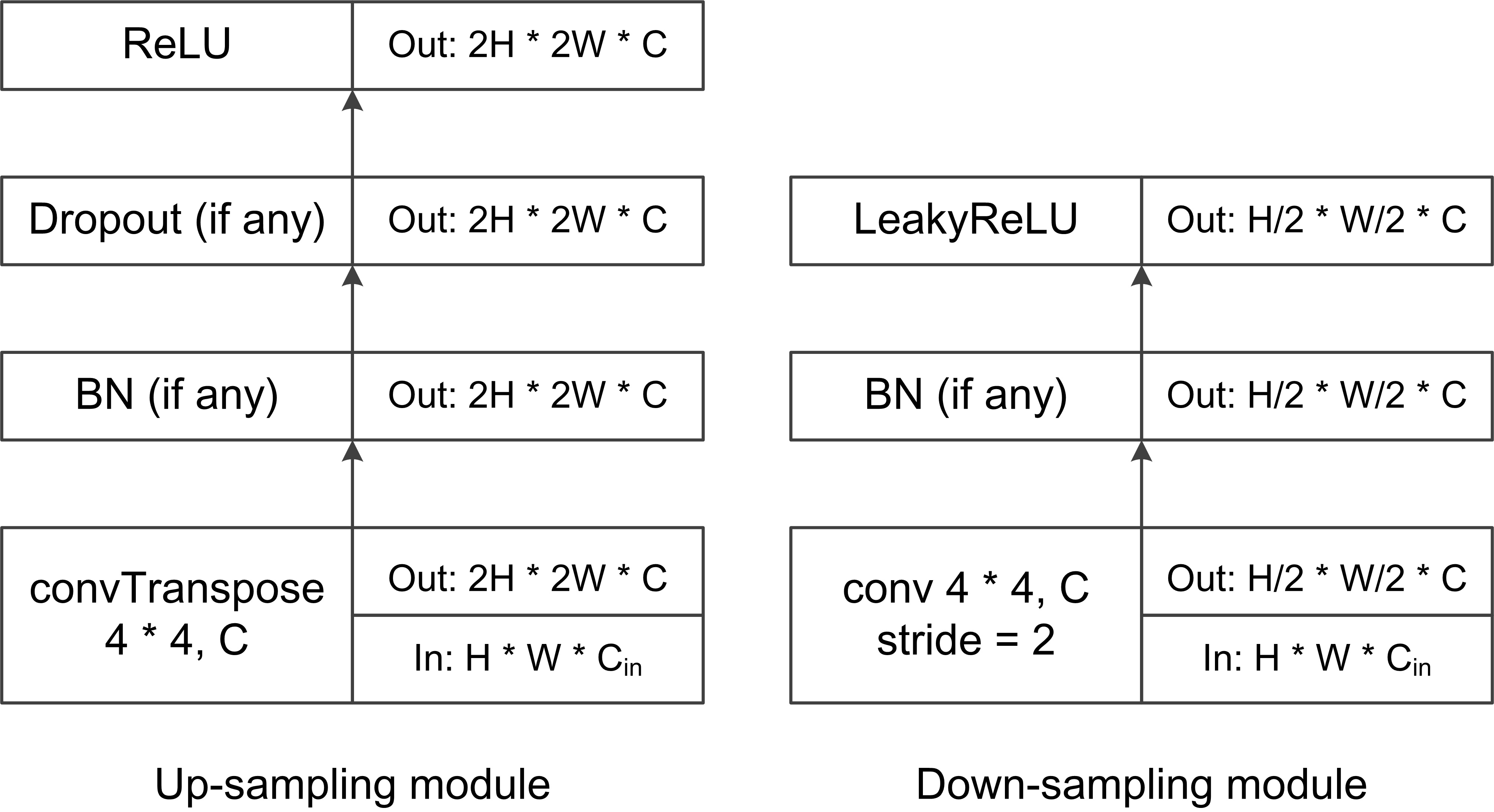}
\caption{Structural details for up-sampling/down-sampling modules. Here, ``conv 4*4, C'' means to apply C kernels with a size of 4$\times$4 to the input for ordinary convolution. ``stride = 2'' makes the output size become (H/2, W/2), where (H, W) is the input size (ignoring the number of channels). ``convTranspose 4*4, C'' means to apply C kernels with a size of 4$\times$4 to the input for transposed convolution \cite{trans:conv}, making the output size become (2H, 2W). Notice that, ``cropping'' and ``max pooling'' \cite{dnn:unet} are not used.
}
\end{figure}

\begin{figure}[!t]
\centering
\includegraphics[width=3.2in]{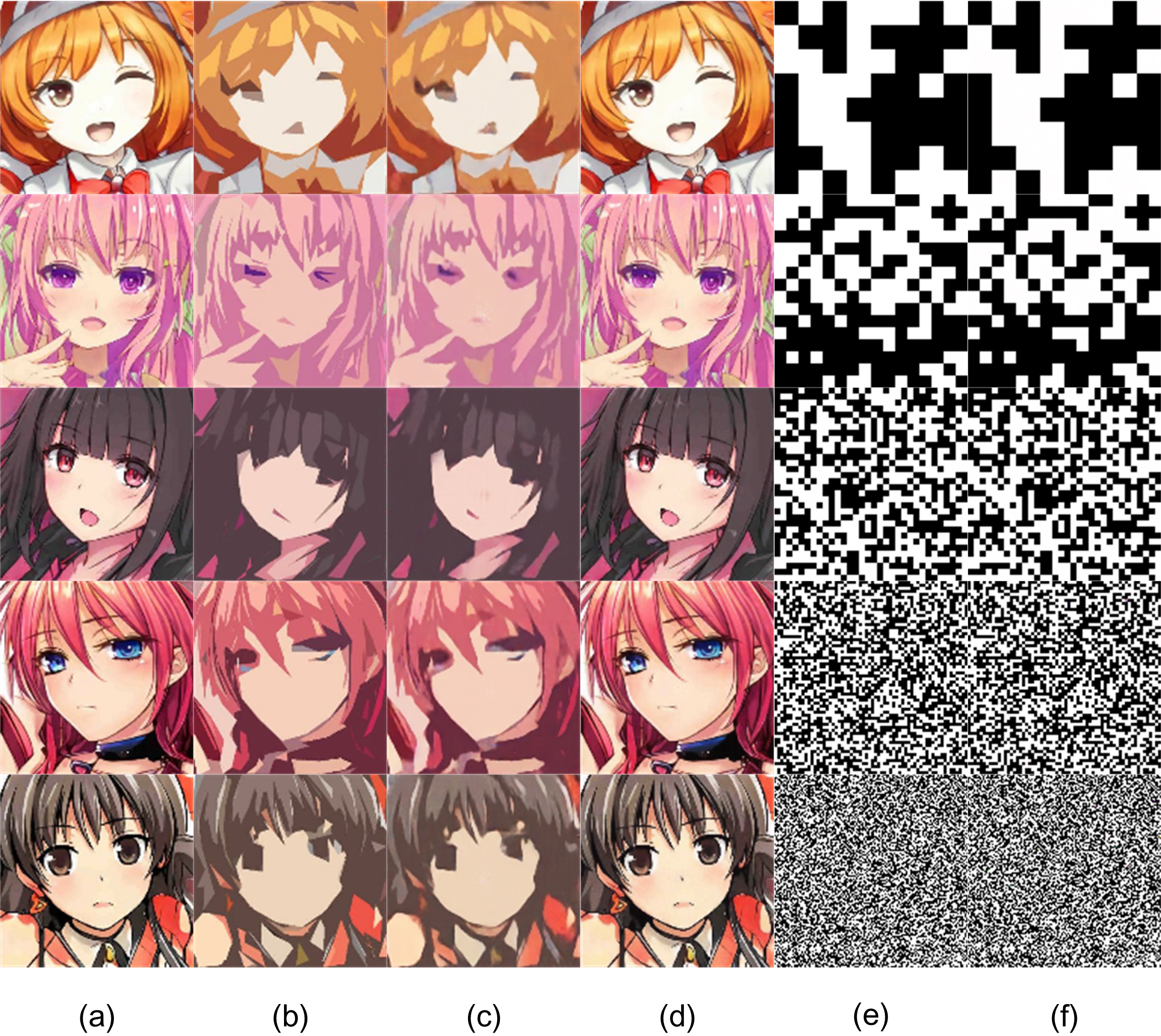}
\caption{Examples: (a) cover image/to-be-transferred image, (b) ground-truth image (style transfer), (c) style transferred image, (d) stego image, (e) secret data, (f) recovered secret data without error. The ALs of secret data are $8^2$, $16^2$, $32^2$, $64^2$ and $128^2$ respectively from top to bottom.}
\end{figure}

\begin{figure}[!t]
\centering
\includegraphics[width=3in]{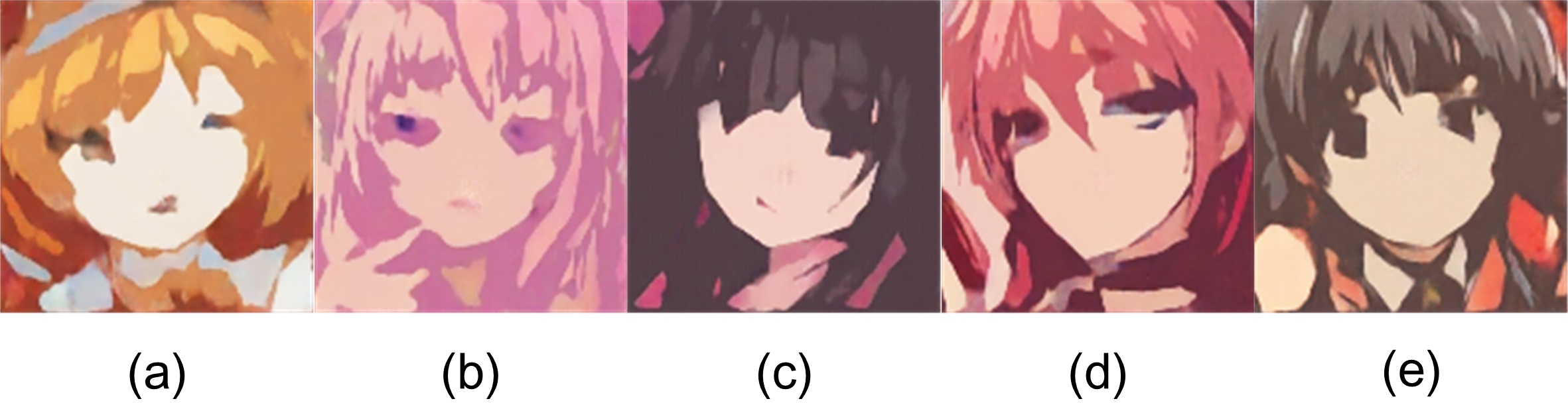}
\caption{The style transferred images for the stego images shown in Fig. 4 (d) due to different ALs: (a) $8^2$, (b) $16^2$, (c) $32^2$, (d) $64^2$ and (e) $128^2$.}
\end{figure}

\subsection{Loss Function}
For style transfer, given $\textbf{x}$ and $\textbf{y}$, the style transferred image $\textbf{z} = F_w(\textbf{x}, \textbf{y})$ should be visually close to the ground-truth image $\textbf{z}_g$. To this end, the loss function is defined as:
\begin{equation}
L_1 = \mathbb{E}_{\textbf{x},\textbf{y},\textbf{z}_g}||F_w(\textbf{x}, \textbf{y}) - \textbf{z}_g||_1^1 / M,
\end{equation}
where $M$ is the number of elements in $\textbf{z}_g$, $||*||_1^1$ corresponds to L1-norm. In Fig. 1 (c), given the cover image \textbf{c} and the secret data \textbf{m}, $F_w$ outputs the stego image \textbf{s}. On the one hand, \textbf{s} should be visually close to \textbf{c}. On the other hand, $F_w$ should be able to fully extract $\textbf{m}$ from \textbf{s} by using the trigger signal \textbf{r} in Fig. 1 (d). We therefore define the loss as:
\begin{equation}
\begin{split}
L_2 = \mathbb{E}_{\textbf{c}, \textbf{m}, \textbf{r}}||F_w(\textbf{c}, &\textbf{m}) - \textbf{c}||_1^1 / M\\
+& ~ \alpha_1\cdot ||F_w(F_w(\textbf{c}, \textbf{m}), \textbf{r}) - \textbf{m}||_1^1 / M,
\end{split}
\end{equation}
where $\alpha_1$ is a tunable parameter. Therefore, the entire loss function for training $F_w$ is:
\begin{equation}
L = L_1 + \alpha_2\cdot L_2,
\end{equation}
where $\alpha_2$ is also tunable. By
default, we use $\alpha_1 = \alpha_2 = 1$ because the three tasks are equally important in this paper.

\emph{Remark:} There are two reasons for explaining why $F_w$ is able to accomplish three different tasks. First, DNNs are known to have many redundant parameters \cite{DL:book}, implying that the undeveloped generalization ability of DNNs can be further exploited to realize other tasks. This can be done by training a joint loss function, which has been utilized by the proposed method (refer to Eq. (3)) and existing works, e.g., Uchida \emph{et al.} \cite{Japan:DNNWatermarking} propose to train a DNN by optimizing a joint loss so that the DNN not only performs well on its original task, but also carries a watermark. Second, different tasks take the significantly different images as input, which can be learned by $F_w$. Clearly, in Fig. (a, c, d), while the images $\{\textbf{x}, \textbf{c}, \textbf{s}\}$ can be the same as each other, the images $\{\textbf{y}, \textbf{m}, \textbf{r}\}$ are significantly different from each other, i.e., $\textbf{y}$ is a meaningful color image, $\textbf{m}$ is a random binary image and $\textbf{r}$ is a fully black image. During training, $F_w$ easily recognizes the difference and accordingly can perform different tasks. The feasibility of using different input images for realizing different tasks has been verified by the existing works, e.g., a secret watermark can be embedded into a DNN by inserting a trigger pattern to a sequence of input images \cite{backdoor:watermarking}. 

\begin{figure}[!t]
\centering
\includegraphics[width=3in]{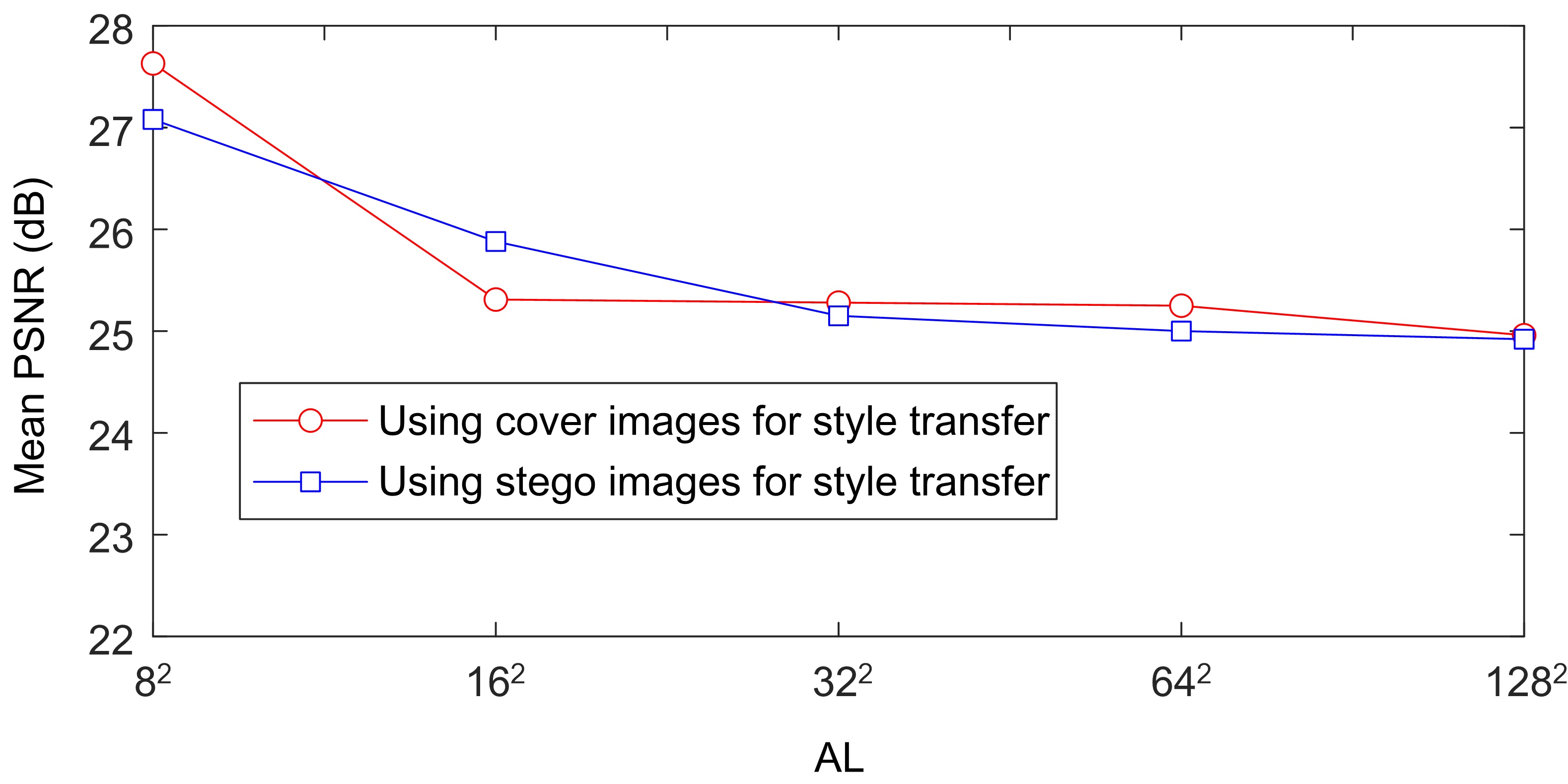}
\caption{The mean PSNRs for style transfer with cover/stego images.}
\end{figure}

\begin{figure}[!t]
\centering
\includegraphics[width=3in]{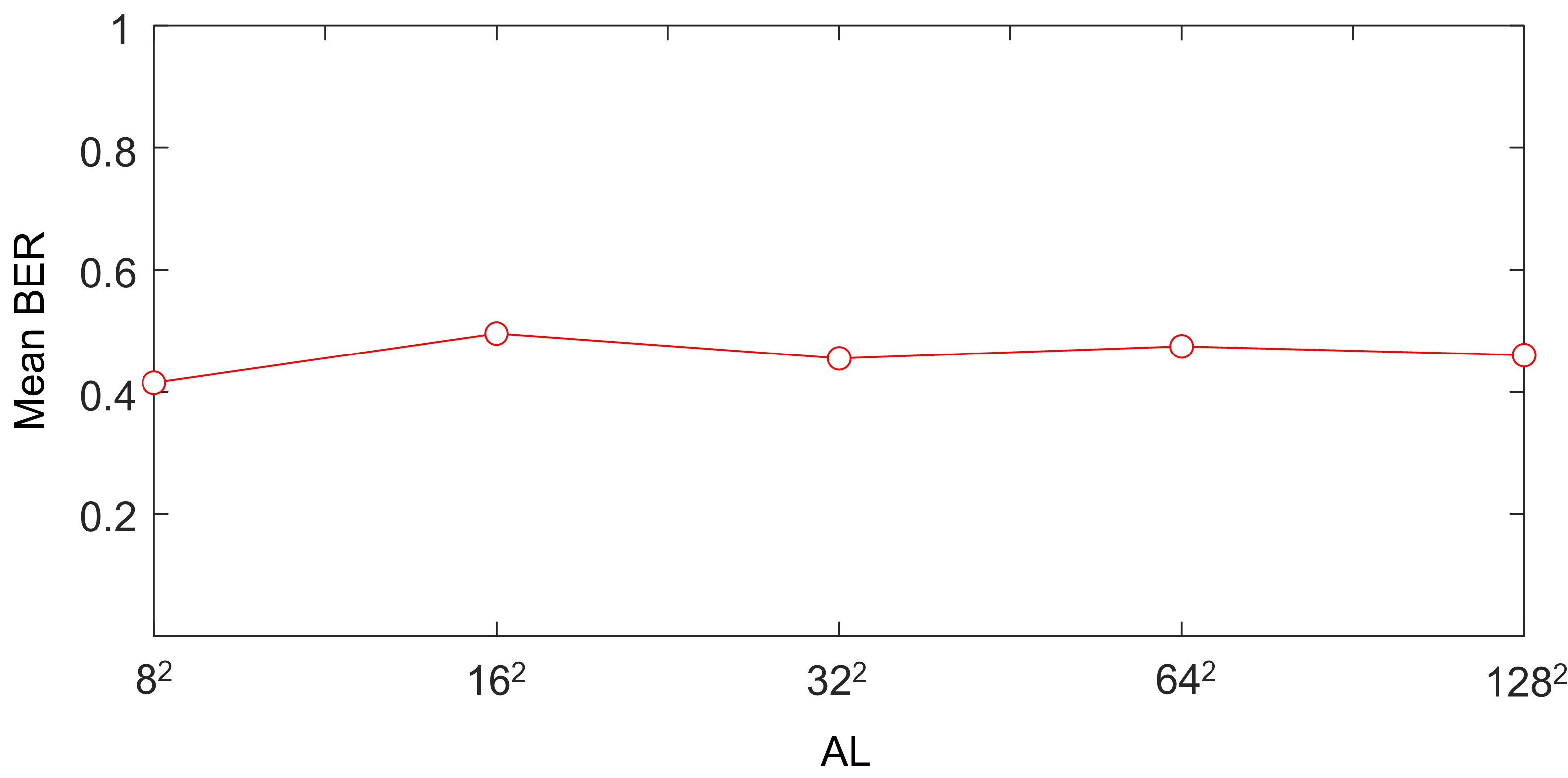}
\caption{The mean BERs for data extraction with a random trigger signal.}
\end{figure}

\section{Experimental Results and Analysis}
\subsection{Setup}
The Danbooru2019 dataset\footnote{\url{https://www.gwern.net/Danbooru2019}} was used for experiments. We randomly selected 12,000, 2,000 and 5,000 images out from the dataset for model training, model validation and model testing. The ADAM optimizer \cite{adam:paper} was used for parameter optimization, where the learning rate was $l_r = 0.0001$ and two hyper-parameters were $\beta_1 = 0.9$ and $\beta_2 = 0.999$. Our implementation used TensorFlow and cuDNN, trained on a single TITAN RTX GPU (24 GB). The batch size was set to 1 due to the limited computational resource. The other default parameters were provided by the TensorFlow platform. 

During model training, for each batch, the DNN should accomplish three tasks. The DNN parameters were updated after the three tasks were all completed. And, we empirically used $\alpha_1 = 1$ and $\alpha_2 = 1$ for the loss function. The original secret information to be embedded was randomly generated as a binary string. For style transfer, we used Photoshop\footnote{\url{https://www.adobe.com/products/photoshop.html}} with default parameter settings to create the ground-truth images for the style transferred images since there has no unified standard method for us to generate the required ground-truth images. All the ground-truth images have the same style to the style image $\textbf{y}$ in Fig. 1 (a, b). The style image was fixed in advance, i.e., all the to-be-transferred images used the same style image. We will show that the style transferred images generated by other methods can be used as the ground-truth images, indicating that our method does not rely on any specific ground-truth generation method.

\begin{table}[!t]
\renewcommand{\arraystretch}{1.1}
\caption{Task difference between the three DNNs $F_w$, $F_{w_1}$ and $F_{w_2}$.}
\centering
\begin{tabular}{c|ccc}
\hline\hline
DNN & Style Transfer & Data Embedding & Data Extraction\\
\hline
$F_w$ & $\surd$ & $\surd$ & $\surd$\\
\hline
$F_{w_1}$ & $\surd$ & $\times$ & $\times$\\
\hline
$F_{w_2}$ & $\times$ & $\surd$ & $\surd$\\
\hline\hline
\end{tabular}
\end{table}

\begin{table}[!t]
\renewcommand{\arraystretch}{1.1}
\caption{The mean PSNR (dB) and mean SSIM (on 5000 test images).}
\centering
\begin{tabular}{c|cccccc}
\hline\hline
DNN/AL & $0$ & $8^2$ & $16^2$ & $32^2$ & $64^2$ & $128^2$\\
\hline
$F_{w_1}$: PSNR & 28.13 & $-$ & $-$ & $-$ & $-$ & $-$\\
$F_{w_1}$: SSIM & 0.865 & $-$ & $-$ & $-$ & $-$ & $-$\\
\hline
$F_w$: PSNR & $-$ & 27.63 & 25.31 & 25.28 & 25.25 & 24.96\\
$F_w$: SSIM & $-$ & 0.860 & 0.800 & 0.800 & 0.796 & 0.776\\
\hline\hline
\end{tabular}
\end{table}

\begin{table}[!t]
\renewcommand{\arraystretch}{1.1}
\caption{The mean PSNR (dB) and mean BER (on 5000 test images).}
\centering
\begin{tabular}{c|ccccc}
\hline\hline
DNN/AL & $8^2$ & $16^2$ & $32^2$ & $64^2$ & $128^2$\\
\hline
$F_{w_2}$: PSNR & 38.29 & 37.36 & 37.18 & 36.31 & 34.83\\
$F_{w_2}$: BER & 0 & 0 & 0 & $5.28\times 10^{-5}$ & $4.10\times 10^{-4}$\\
\hline
$F_w$: PSNR & 37.00 & 36.68 & 36.49 & 36.20 & 31.45\\
$F_w$: BER & 0 & 0 & 0 & $7.48\times 10^{-5}$ & $4.62\times 10^{-4}$\\
\hline\hline
\end{tabular}
\end{table}

\begin{table}[!t]
\renewcommand{\arraystretch}{1.1}
\caption{Classification accuracy with the steganalysis method in \cite{xu:net}.}
\centering
\begin{tabular}{c|ccccc}
\hline\hline
DNN/AL & $8^2$ & $16^2$ & $32^2$ & $64^2$ & $128^2$\\
\hline
$F_{w_2}$ & 0.906 & 0.925 & 0.939 & 0.975 & 0.995\\
\hline
$F_w$ & 0.913 & 0.9375 & 0.941 & 0.981 & 0.998\\
\hline\hline
\end{tabular}
\end{table}

\subsection{Qualitative Results and Image Style Transfer}
Fig. 4 shows some examples of applying $F_w$ to style transfer, data embedding and data extraction. It can be observed that all the generated images have good visual quality. Moreover, the secret data can be reliably recovered, which has verified the feasibility of the proposed method. In Fig. 1, when the data receiver is being monitored by the behavior monitor B, he should use the stego image for style transfer. To verify the feasibility, we used the stego images shown in Fig. 4 (d) for style transfer. As shown in Fig. 5, the style transferred images are visually close to the corresponding ground-truth images shown in Fig. 4 (b). We also used all images in the testing dataset and their stego versions for style transfer. The mean peak signal-to-noise ratio (PSNR) was used to measure the difference between the style transferred images and the ground-truth images. Fig. 6 shows the experimental results. It is inferred from Fig. 6 that there is not significant performance difference between the cover images and the stego images, which indicates that the data receiver can well conceal the data extraction tool by using a just received stego image for style transfer when he is being monitored. Fig. 7 shows the bit error rates (BERs) by using a random trigger signal for data extraction. It can be seen that the BERs are close to 0.5, which means that one cannot extract secret data if he does not know the trigger signal in practice.

For fair comparison, we trained two new DNNs $F_{w_1}$ and $F_{w_2}$, where the former was trained with Eq. (1) and the latter was trained with Eq. (2). By comparing $F_w$ with $F_{w_1}$, we can evaluate the performance of $F_w$ on style transfer. By comparing $F_w$ with $F_{w_2}$, we can evaluate the performance of $F_w$ on data hiding. $F_w$, $F_{w_1}$ and $F_{w_2}$ have the same network architecture. The difference is that they were trained with different loss functions. Therefore, they have different task abilities, which can be found from Table 1. Namely, $F_w$ can complete all tasks. $F_{w_1}$ can complete style transfer only. $F_{w_2}$ can complete data embedding and data extraction. 

$F_{w_1}$ is expected to outperform $F_w$ for style transfer since $F_{w_1}$ only learned style transfer. Similarly, $F_{w_2}$ is expected to provide the better performance than $F_w$ for data embedding and data extraction. Table 2 has shown the mean PSNRs and the mean structural similarities (SSIMs) \cite{ssim:paper} due to different settings. When the AL increases, both the mean PSNR and the mean SSIM will decline. Comparing with $F_{w_1}$, the PSNR degradation and SSIM degradation for $F_w$ are kept very low, which means that DH does not impair style transfer.

\subsection{Payload-Distortion Performance and Robustness}
As mentioned above, we used PSNR to evaluate the visual quality of a stego image and BER to measure the difference between the original secret data and the reconstructed data. Table 3 shows the experimental results on 5000 test images. It is observed that the mean PSNR will decline as the AL increases. When the AL is no more than $64^2$, all mean PSNRs are above 35 dB. Even the AL equals the total number of pixels, the mean PSNR is higher than 30 dB, indicating that the stego images have satisfactory visual quality.

Though the mean BER is no longer zero for a large AL, its value is quite low. E.g., when the AL is $128^2$, the mean BER is $4.62\times 10^{-4}$, meaning that the total number of errors is $4.62\times 10^{-4}\times 128^2 = 7.57 << 16384~( = 128^2)$ averagely. In case that the data hider has the trigger signal, the data hider can select any cover image that enables the data receiver to perfectly recover secret data from the corresponding stego image. The reason is that the data hider could skip a cover image if the secret data cannot be perfectly reconstructed from the corresponding stego image by using the trigger signal. Actually, no matter the data hider knows the trigger signal or not, error correcting code can be used, which has been suggested by related arts such as \cite{steganoGAN}, \cite{hu:paper}. Therefore, the data embedding capacity for the proposed method can be considered as 1 bpp (bits per pixel). In Table 3, the mean PSNR difference and the mean BER difference between $F_{w_2}$ and $F_w$ are very low. For example, when the AL equals $128^2$, the mean BER difference is $0.52\times 10^{-4}$, which means there are only $0.52\times 10^{-4}\times 128^2 = 0.85$ error bits. Therefore, it can be inferred that HDH can be reliably realized in practice. 


A stego image may be maliciously/unintentionally modified before data extraction, which inspires us to use adversarial training \cite{wu:tcsvt:2020} for improving the robustness. To mimic a realistic scenario, we added Gaussian noise to stego images with $\mu = 0$ and random $\sigma \in \{0.01, 0.05, 0.1, 0.15\}$ for training. The mean BER was obtained to evaluate the robustness. The ALs $64^2$ and $128^2$ were tested as they resulted in a very small number of error bits even no noise was added. Fig. 8 shows the experimental results, from which we can find that training with the noised data can significantly improve the robustness. It indicates that it is desirable to use adversarial training to resist against a specific attack in practice.

\begin{figure}[!t]
\centering
\includegraphics[width=3.3in]{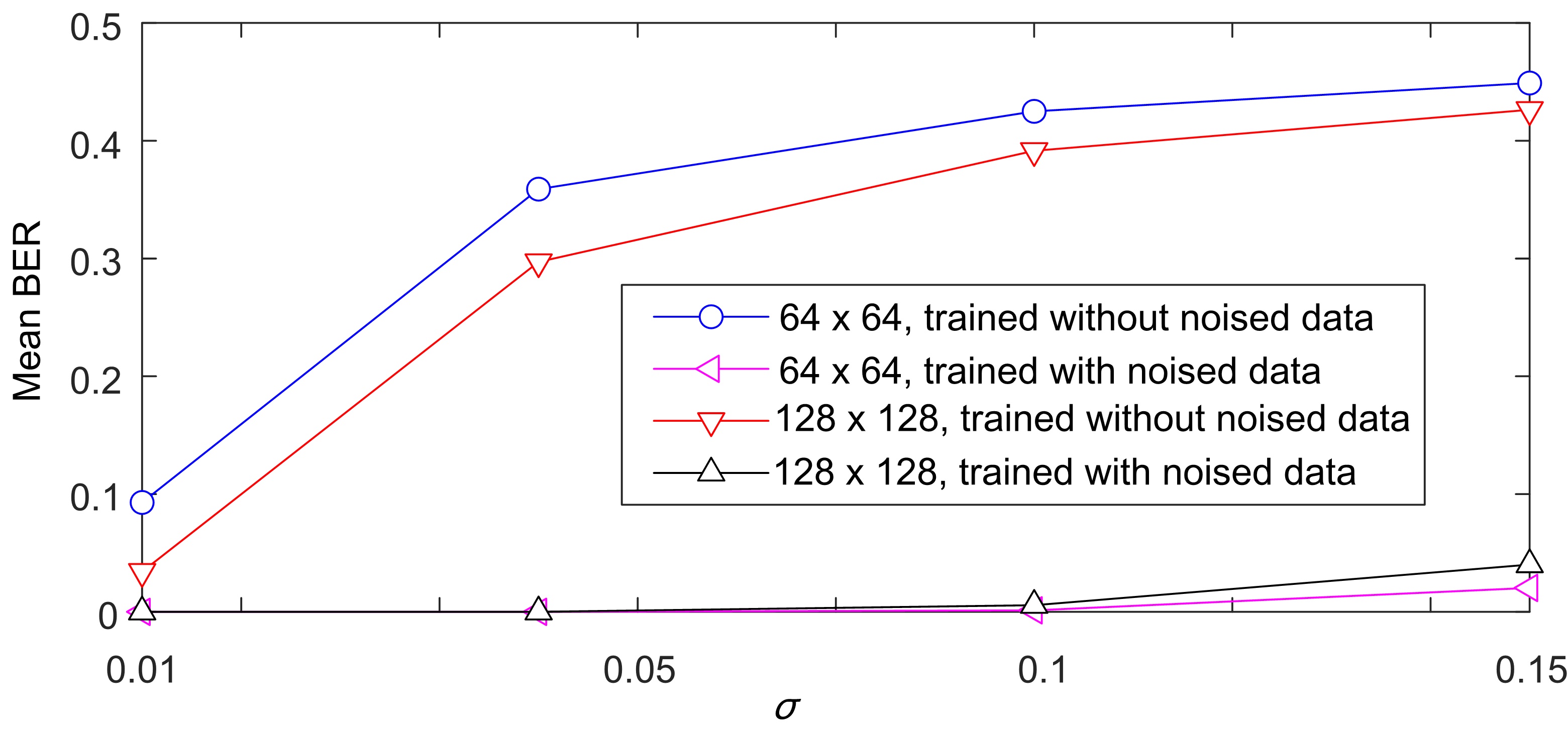}
\caption{The resultant mean BERs with/without noised data for training.}
\end{figure}

\subsection{Security Analysis}
The previous 5,000 test images and their stego versions were used for image steganalysis. The benchmark model XuNet \cite{xu:net} was used for binary classification. We used 80\% images for training (in which 10\% for validation) and 20\% images for testing. The classification accuracy was computed as the ratio of correctly classified samples during testing. Table 4 shows the experimental results. When the AL increases, the classification accuracy increases as well. All the classification accuracy values are high for both $F_{w_2}$ and $F_w$. The reason is that we did not use any adaptive embedding mechanism in the DNN but XuNet is effective in detecting non-adaptive embedding. Designing the adaptive embedding mechanism should take into account the impact caused by style transfer, which is another topic needing further study. Nevertheless, it can be inferred that the difference between $F_{w_2}$ and $F_w$ in terms of the classification accuracy is very low. It indicates that HDH does not impair security when the network architecture has been fixed in advance. Notice that, $F_{w_2}$ has better performance on DH due to the fewer tasks to be learned.

It should be pointed that the above steganalysis experiments rely on the strict assumption that the content steganalyzer has the power to collect a number of cover images and stego images. A DH system is deemed secure (to a certain extent) if it manages to fool the steganalyzer even under such kind of disadvantageous condition. However, from the viewpoint of real-world applications, it is almost impossible for the steganalyzer to collect a number of labeled samples for steganalysis because he does not know the existence of DH tools (including the trigger signal). Otherwise, it will go against our original intention. Therefore, by hiding DH, the system security can be significantly improved compared to previous works that does not conceal the DH tools.

\begin{table}[!t]
\renewcommand{\arraystretch}{1.1}
\centering
\caption{Quantitative results by using the L2-norm loss. The PSNRs, SSIMs and BERs shown in this Table are mean values (on 5000 test images).}
\begin{tabular}{c|c|c|c|c}
\hline\hline
\multicolumn{1}{c|}{\multirow{2}{*}{AL}} &
\multicolumn{2}{c|}{Style Transfer} & \multicolumn{1}{c|}{Data Embedding} & \multicolumn{1}{c}{Data Extraction} \\
\cline{2-5}
& PSNR (dB) & SSIM  &  PSNR (dB) & BER \\
\hline
$16^2$  & 25.60 & 0.814 & 36.63 & 0\\
$64^2$  & 25.17 & 0.793 & 36.08 & $7.97\times 10^{-5}$\\
$128^2$ & 25.01 & 0.770 & 31.94 & $4.58\times 10^{-4}$\\
\hline\hline
\end{tabular}
\end{table}

\subsection{Ablation Study}
In this paper, we use the L1-norm loss function for model training (please refer to Eq. (1) and Eq. (2)). Comparing with the L2-norm loss, the L1-norm loss is more robust against possible outliers in a dataset \cite{image2image:paper}. In most cases, both work very well. In this ablation study, we will show that both can be used for the proposed work since their performance are close to each other. In detail, in experiments, the tested ALs were $16^2$ (low embedding rate), $64^2$ (medium embedding rate) and $128^2$ (high embedding rate). For model training, the L1-norm loss in Eq. (1) and the L1-norm loss in Eq. (2) were all modified with the L2-norm loss. Table 5 shows the experimental results. By comparing Table 5 and Tables 2-3, we can find that there is not significant difference between using L1-norm loss and using L2-norm loss. Therefore, it can be inferred that both losses can be used for the proposed work. It should be also admitted that one may design other efficient loss functions for training $F_w$. For example, it is possible that one can use a more advanced loss function for style transfer (since style transfer has been widely studied in the literature), which is not the main focus of this paper. 

Previously, the ground-truth images used for style transfer were created by Photoshop. To show that the proposed method does not rely on any specific ground-truth generation method, we used the pre-trained DNN proposed by Li \emph{et al.} \cite{styleTransfer:NewPaper} to generate the ground-truth images. The reason for choosing the DNN tool is that the DNN shows superior performance in style transfer and more importantly it also accepts two images as input to generate one output image, which is very convenient for us to create the ground-truth images. As shown in Fig. 9, all the generated images have satisfactory visual quality. Notice that, the style transferred images in Fig. 9 (g) are corresponding to the stego images in Fig. 9 (d), and the style transferred images in Fig. 9 (c) are corresponding to the cover images in Fig. 9 (a). To provide the quantitative results, we used the 5000 test images mentioned in Subsection 3.1 for completing style transfer, data embedding and data extraction. For style transfer, the mean PSNR and the mean SSIM between the above 5000 test images and the corresponding style transferred images were tested. For data embedding, the mean PSNR between the test images and the corresponding stego images was tested. And, the mean BER (between the original secret data and the reconstructed data) was used to evaluate the performance of data extraction. As shown in Table 6, for style transfer, the mean PSNRs are higher than 34 dB and the mean SSIMs are close to 0.87. Moreover, for data embedding, the mean PSNRs are all higher than 37 dB. It indicates that both style transfer and data embedding generate high-quality images. For data extraction, the mean BERs are very low, implying that the secret data can be perfectly recovered in most cases. Therefore, we can conclude that the proposed method has good applicability and will not be subjected to any specific ground-truth generation method. In addition, by comparing Table 6 with Tables 2-3, it seems that it is more desirable to use the well-trained DNN to generate the ground-truth images, which may be due to the reason that a well-trained DNN can generate images more conducive to training $F_w$.

\begin{figure}[!t]
\centering
\includegraphics[width=\linewidth]{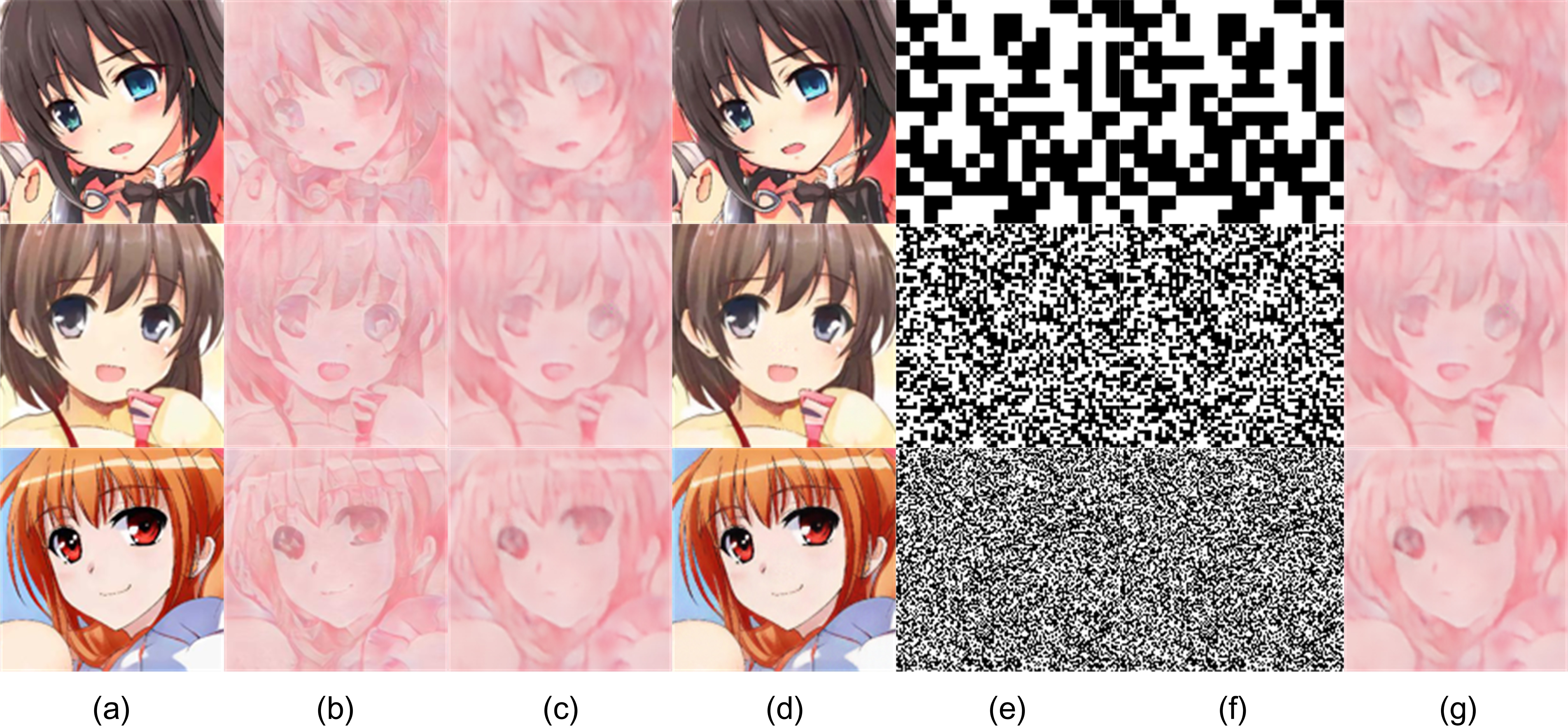}
\caption{Examples: (a) cover image, (b) ground-truth image (used for style transfer), (c) style transferred image (for the cover), (d) stego image, (e) secret data, (f) reconstructed data without error, and (g) style transferred image (for the stego). The ALs are $16^2$, $64^2$ and $128^2$ from top to bottom.}
\end{figure}

\begin{table}[!t]
\renewcommand{\arraystretch}{1.1}
\centering
\caption{Quantitative results by using a pre-trained DNN to generate the ground-truth images for the style transfer task. The PSNRs, SSIMs and BERs shown in this Table are mean values (on 5000 test images).}
\begin{tabular}{c|c|c|c|c}
\hline\hline
\multicolumn{1}{c|}{\multirow{2}{*}{AL}} &
\multicolumn{2}{c|}{Style Transfer} & \multicolumn{1}{c|}{Data Embedding} & \multicolumn{1}{c}{Data Extraction} \\
\cline{2-5}
& PSNR (dB) & SSIM  &  PSNR (dB) & BER \\
\hline
$16^2$  & 35.04 & 0.876 & 41.12 & 0\\
$64^2$  & 34.99 & 0.875 & 41.05 & $4.60\times 10^{-5}$\\
$128^2$ & 34.73 & 0.869 & 37.42 & $2.66\times 10^{-4}$\\
\hline\hline
\end{tabular}
\end{table}

\subsection{More Comparisons}
In this subsection, we further compare the proposed HDH method with related works for evaluation even though the related works were not originally designed for ``hiding data hiding'', but rather ``data hiding''. We here choose the novel method proposed by Baluja \cite{DH:TPAMI} (an extension version of the conference paper \cite{TPAMIpaper:conferenceVersion}) for fair comparison. The reason is that the method enables us to hide a secret image into a cover image to generate a stego image, and reconstruct the secret image only with the stego image, which closely matches the proposed method and is quite suitable for comparison. 

We used the dataset and parameter settings mentioned in Subsection 3.1 for simulation. It is noted that the secret image in our experiments was set to a randomly generated binary image, rather than a color image mentioned in \cite{DH:TPAMI}. For each tested AL, we determined the mean PSNR between the stego images and their cover versions to evaluate the visual quality of the stego images. And, we determined the mean BER to evaluate the performance of data extraction. Table 7 shows the experimental results. By comparing Table 7 with Table 3, it can be inferred that the visual quality of the stego images generated by \cite{DH:TPAMI} outperforms the proposed method. We explain this reasonable phenomenon by analyzing the significant difference between the method in \cite{DH:TPAMI} and the proposed method. In detail, the method in \cite{DH:TPAMI} combines three independent sub-networks to realize data hiding. Even though the three sub-networks can be well trained together, each sub-network has its own goal and will not impair the other sub-networks. In other words, in \cite{DH:TPAMI}, a sub-network well realizes data embedding and another sub-network well realizes data extraction. However, for the proposed method, only one network is used. The network should accomplish three tasks, which will significantly increase the difficulty of network training and thus reduce the generalization ability. Moreover, the network architecture reported in this paper may be not optimal. Therefore, though the proposed method already achieves satisfactory performance, the stego images generated by \cite{DH:TPAMI} have the relatively better quality. 

\begin{table}[!t]
\renewcommand{\arraystretch}{1.1}
\centering
\caption{Quantitative results for the method in \cite{DH:TPAMI}. The PSNRs and BERs shown in this Table are mean values (on 5000 test images).}
\begin{tabular}{c|c|c|c|c}
\hline\hline
\multicolumn{1}{c|}{\multirow{2}{*}{AL}} &
\multicolumn{2}{c|}{Style Transfer} & \multicolumn{1}{c|}{Data Embedding} & \multicolumn{1}{c}{Data Extraction} \\
\cline{2-5}
& PSNR (dB) & SSIM  &  PSNR (dB) & BER \\
\hline
$16^2$  & $-$ & $-$ & 38.50 & 0\\
$64^2$  & $-$ & $-$ & 37.79 & $1.74\times 10^{-6}$\\
$128^2$ & $-$ & $-$ & 35.12 & $4.03\times 10^{-5}$\\
\hline\hline
\end{tabular}
\end{table}

\begin{table}[!t]
\renewcommand{\arraystretch}{1.2}
\caption{Classification accuracy for detecting the method in \cite{DH:TPAMI} with XuNet.}
\centering
\begin{tabular}{c|ccc}
\hline\hline
DNN/AL & $16^2$ & $64^2$ & $128^2$\\
\hline
\cite{DH:TPAMI} & 0.935 & 0.966 & 0.990\\
\hline\hline
\end{tabular}
\end{table}

Similarly, it can be seen from Table 7 that the method in \cite{DH:TPAMI} inevitably produces error-bits even though the BER is extremely low. Actually, the existing works such as \cite{steganoGAN} have indicated that the ability to recover a hidden bit is heavily dependent on the DNN model and the cover image, as well as the secret data itself. In other words, it is often the case that a DNN cannot perfectly reconstruct secret data (for a relatively high embedding rate) without any error correcting mechanism. The methods in \cite{Hidden:Fei-Fei-Li}, \cite{steganoGAN}, \cite{LFM:CVPR}, \cite{ABDH:AAAI:Paper}, \cite{DH:TPAMI}, \cite{hu:paper}, \cite{wu:tcsvt:2020}, \cite{Hayes:paper} are all in line with the above analysis. Therefore, as suggested in \cite{steganoGAN} and \cite{hu:paper}, error correcting codes (e.g., Reed Solomon codes \cite{RScode}) can be utilized for data extraction. It is worth mentioning that, though the method in \cite{Liu:ICAIS} enables secret data to be perfectly reconstructed, it requires the data hider to carefully choose the cover image. Moreover, it was designed for classification based DNNs (rather than generative DNNs), which has a very low embedding capacity.

Furthermore, we used the aforementioned model XuNet for detecting the stego images generated by the method in \cite{DH:TPAMI} to evaluate the system security. The parameter settings were the same as that described in Subsection 3.4. Table 8 shows the experimental results, from which we can infer that the method in \cite{DH:TPAMI} cannot resist against XuNet either though it is slightly better than the proposed method by comparing Table 8 with Table 4. This indicates that generative DH (which means to use a DNN to directly generate a stego image) is indeed not robust to advanced steganalysis systems at the moment, which has been reported in previous works \cite{DH:TPAMI}, \cite{hu:paper}, \cite{TPAMIpaper:conferenceVersion}. Comparing with cost based DH methods that assign different costs to individual pixels, generative DH methods often assume that the costs of modifying individual pixels are equal to each other. A significant evidence is that the loss function designed for generative DH (e.g., L1-norm or L2-norm loss) implies that the unit loss of modifying a single pixel is constant, which cannot resist against advanced steganalysis systems. It leads to an immediate direction for future work, for which designing a good loss and adversarial training will be helpful.

\section{Conclusion and Discussion}
In this work, we propose a novel DNN based HDH method, which camouflages the DH tools as a DNN with an ordinary task (i.e., style transfer). The proposed HDH method trains the DNN by optimizing a joint loss so that the trained DNN not only can transfer the style of an image to the target one, but also can embed secret data into a cover image or extract secret data from a stego image. In this way, when the data hider (or the data receiver) is being monitored, he uses the DNN to complete style transfer. Otherwise, the DNN is used for data embedding (or data extraction). In other words, the DH tools are well concealed by style transfer. 

Extensive experiments have been conducted to evaluate the proposed method. The experimental results indicate that HDH can be successfully realized and would not impair the ordinary task. Though the BER is no longer zero for a larger AL, its value is kept extremely low, implying that the secret data could be perfectly recovered by error correcting code. It should be also admitted that, for the existing generative DH methods (including the proposed method), even though the stego image has satisfactory visual quality, its ability to resist against steganalysis is not satisfactory, which requires further study. It will be interesting to extend the adaptive data-embedding mechanism designed for cost based DH to generative DH to further enhance the security performance. We hope this first attempt can inspire more advanced works.

\ifCLASSOPTIONcaptionsoff
  \newpage
\fi

\end{document}